\documentclass[conference]{IEEEtran}
\IEEEoverridecommandlockouts

\usepackage{labsedc}
\usepackage{extracommands}
\usepackage{amsmath,amssymb}

\usepackage{hyperref}

\usepackage{caption}
\usepackage{subcaption}

\usepackage{booktabs}
\usepackage{longtable}
\usepackage{supertabular}
\usepackage{multirow}

\usepackage{balance}

\def\BibTeX{{\rm B\kern-.05em{\sc i\kern-.025em b}\kern-.08em
    T\kern-.1667em\lower.7ex\hbox{E}\kern-.125emX}}

\newtheorem{definsec}{Definition}
\begin{document}

\title{Cross-coverage testing of functionally equivalent programs\\
}

\author{\IEEEauthorblockN{Antonia Bertolino}
\IEEEauthorblockA{\textit{ISTI-CNR} \\
Pisa, Italy\\
antonia.bertolino@isti.cnr.it}
\and
\IEEEauthorblockN{Guglielmo De Angelis}
\IEEEauthorblockA{\textit{IASI-CNR} \\
Rome, Italy \\
guglielmo.deangelis@iasi.cnr.it}
\and
\IEEEauthorblockN{Felicita Di Giandomenico}
\IEEEauthorblockA{\textit{ISTI-CNR} \\
Pisa, Italy\\
felicita.digiandomenico@isti.cnr.it}
\and
\IEEEauthorblockN{Francesca Lonetti}
\IEEEauthorblockA{\textit{ISTI-CNR} \\
Pisa, Italy\\
francesca.lonetti@isti.cnr.it}
}

%\IEEEoverridecommandlockouts
\IEEEpubid{\begin{minipage}{\textwidth}\ \\ \\ \\ [12pt] \centering
  Peer-reviewed version accepted for publication at the 4th ACM/IEEE International Conference on Automation of Software Test (AST 2023), May 15--16, 2023, Melbourne, AU
\end{minipage}} 
\maketitle
\IEEEpubidadjcol

\begin{abstract}
Cross-coverage of a program $P$ refers to the test coverage measured over a different program $Q$ that is functionally equivalent to $P$. The novel concept of cross-coverage can find useful applications in the test of redundant software. We apply here cross-coverage for test suite augmentation and show that additional test cases generated from the coverage of an equivalent program, referred to as cross tests, can increase the coverage of a program in more effective way than a random baseline. We also observe that -contrary to traditional coverage testing- cross coverage could help finding (artificially created) missing functionality faults.
\end{abstract}


\section{Introduction}
\label{sec:intro}

Code coverage testing has been around for decades. Many  criteria of increasing sophistication have been defined in hundred academic publications~\cite{zhu1997software,DBLP:journals/cj/YangLW09,DBLP:journals/tosem/MirandaB20}. On the other side, basic (and less costly) techniques as statement or branch coverage are routinely used in industry~\cite{nativ2001cost,ivankovic2019code}, seconding Beizer's early recommendation that covering all branches should be \textit{the minimum mandatory test requirement}~(\cite{Beizer}, p.~75).

Although their actual effectiveness in failure detection is  debated~\cite{inozemtseva2014coverage,DBLP:journals/tse/GaySWH15}, coverage testing techniques remain  attractive because they promote a systematic and measurable test process~\cite{mockus2009test,fischer2022itest} and allow for automated test generation~\cite{Anand}. One shared argument between detractors and supporters of code-based testing is that achieving full coverage should not be by itself the goal for testing; rather, measuring coverage of otherwise conceived tests (e.g., functional test cases) can be invaluable to identify parts of the program that have not been executed and hence could hide  faults that would otherwise go unnoticed. In fact, one well known limit of code-based test techniques is that by construction they cannot detect missing functionality problems, because they can only test what is in the code.

Basically, then, a schematic \textit{how-to} for code-based testing would be \textit{(i)} to monitor the coverage percentage achieved on the software under test (SUT) by the current test suite, and, until this is not satisfactory, \textit{(ii)} to identify (manually or automatically)  additional test cases (i.e., test suite augmentation) that can execute the yet untested parts of the code. In such a process, it is  understood that the SUT code is available for instrumentation, and it is also tacitly assumed that additional tests not contributing to increase coverage are less, if at all, useful.

In the current literature, the above-described process concerns one single implementation at a time, i.e., the SUT is one program (unit, method, class, \dots ) implementing a given function: this SUT is monitored for measuring coverage, and the additional test cases are identified looking at which parts of the SUT have not been exercised.
In this work, we introduce the novel concept of \textit{cross-coverage testing} for the case that multiple implementations of a same function have to be tested. Intuitively, this term
refers to testing one program using the tests obtained for covering the code of
a different, but functionally equivalent, implementation than the one under exam. Said in different words, we investigate if and how the well-established notion of code-coverage testing could be usefully deployed to the case of redundant software.

In literature, software redundancy has been leveraged in different contexts and with differing aims. For instance, multi-version programs have been used for decades in safety-critical applications as a means to tolerate residual faults~\cite{verhaegen2010fault,voges2012software}. In web service applications it can often be the case that multiple services exposing a same API are found and before one is selected they need to be tested to validate their functionality~\cite{ngan2013semantic}. Similarly, in the many available open source libraries several different functions can be retrieved that implement a same utility, e.g., a mathematical function. In general, we consider here the case that two or more programs\footnote{We will use the generic term of ``program'' throughout, whereby the notion of cross-coverage can be adapted for the testing of methods, classes or whole systems.} exist that are functionally equivalent.

The overarching question we address is: given two functionally equivalent programs $P$ and $Q$, can we leverage the information relative to code coverage of $Q$ to improve the testing of $P$ (or the vice versa)? In particular, with reference to the above schematic \textit{how-to} process, if we derive additional test cases to increase, say,  the coverage of $Q$, do such additional test cases, which we call \textit{cross tests}, provide useful insights if used instead to test $P$? While the problem of testing functionally equivalent programs has been investigated from different perspectives, as we will discuss in the Related Work section, we are not aware of any previous work that has addressed the above questions.

In this paper we define the concept of cross-coverage of functionally equivalent programs, and empirically study its application in test suite augmentation. We conducted an evaluation over a benchmark of 336 programs collected into 140 functionally equivalent groups~\cite{higo2022}. To assess the potential thoroughness of cross-coverage test suite augmentation, we first tested all programs in a group using a common test suite derived by the well-known tool Randoop\footnote{https://randoop.github.io/randoop/}. Then, on each program in turn, we measured the (own)
coverage increase achieved by the additional cross tests derived from each equivalent program in the same group. We  observed an average increase of $\sim26\%$ in statement coverage and of $\sim33\%$ in branch coverage.
The gains have been also confirmed when we restricted the study to a subset of programs from $7$ groups for which we automatically found additional random test cases to compare as a baseline against the cross tests: for them the cross-tests achieve a coverage increase that is on average $\sim21\%$ higher for statement and $\sim22\%$ higher for branch than the increase obtained by the random augmentation with an equal number of  tests.
We also conducted an empirical study for assessing the efficacy of cross coverage in finding missing functionalities: i.e., we deleted some code parts and used the cross tests to see whether the tests were able to raise a failure. Over $169$ programs where the injected fault remained undetected, our results show that cross-coverage test suite augmentation found $\sim73\%$ of them. Thus, we could automatically conceive test cases covering  missing functionality faults for a program $P$ from the throughout coverage of a functionally equivalent program $Q$.

The paper is structured as follows. In the next section we present two motivating examples. Then, in Section~\ref{sec:approach} we propose a formal definition of cross-coverage testing and related concepts. In Section~\ref{sec:studyDesign} we introduce the empirical evaluation study, whose results are then described in  Section~\ref{sec:studyResults}, while in Section~\ref{sec:threats} we discuss some aspects threatening the validity of this study. We contrast our work against existing literature in Section~\ref{sec:related}, and finally draw conclusions in Section~\ref{sec:conclusions}.

\section{Motivation}
\label{sec:motivation}

In simplified words, what we investigate here is whether the test cases that are generated based on the code-coverage of a given program can be leveraged to also test one or more other programs that are (supposed to be)  functionally equivalent to the former. Before formally defining in the next section the notion of using coverage measures across functionally equivalent programs, we discuss here how and why this notion could be useful using a couple of examples.

\begin{figure}
	\centering
	 \includegraphics[width=0.9\columnwidth]{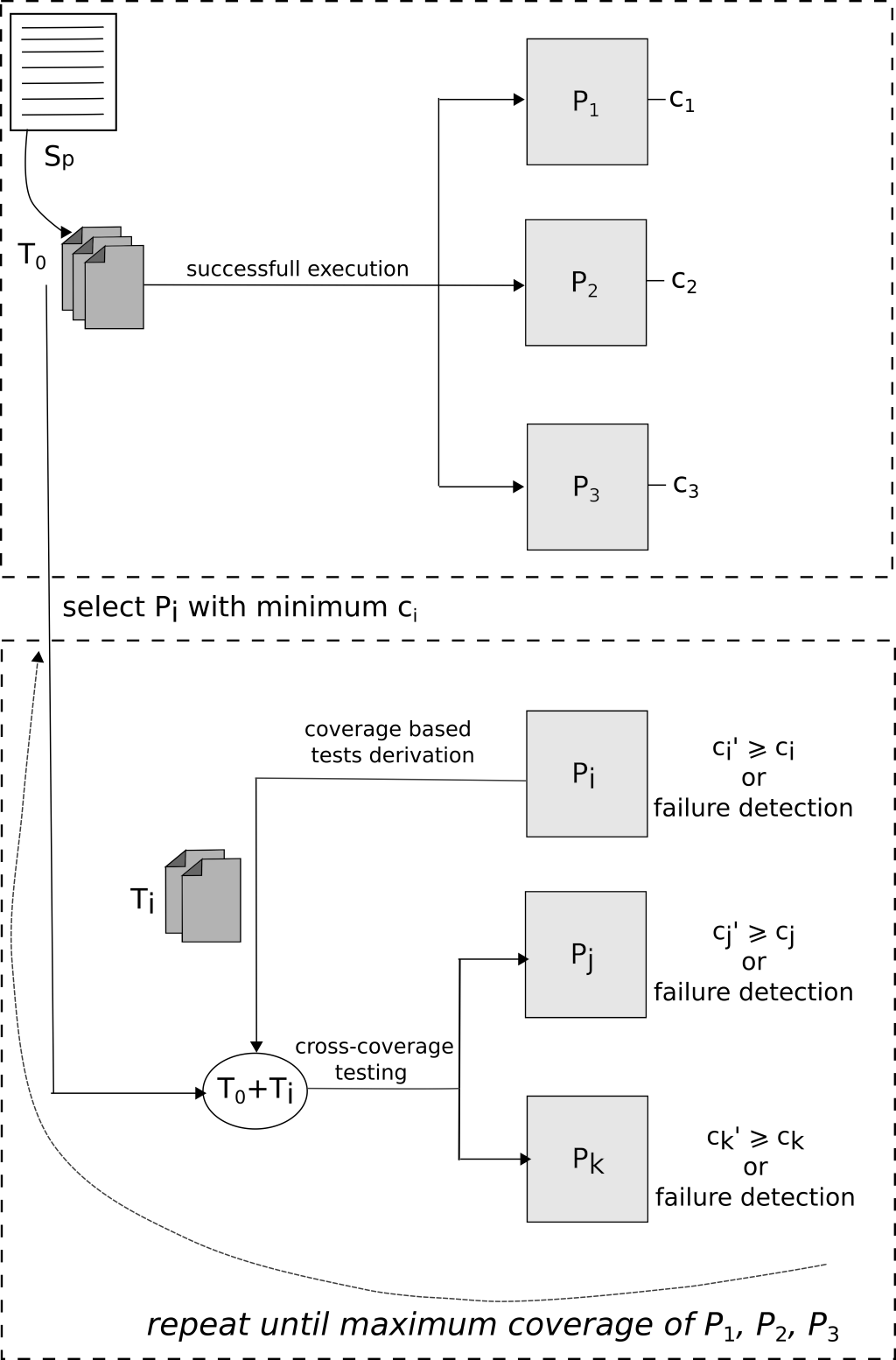}
	\caption{\label{fig:firstExample} \textit{Cross-coverage testing} for \emph{N-version} programming.}
\end{figure}

The very first motivating example (illustrated in Figure \ref{fig:firstExample}) that we can think of belongs to the realm of N-version programming, in which several versions of a same program are executed in parallel for ensuring fault tolerance~\cite{avizienis1985n}. To mitigate the occurrence of common mode failures, it is important to ensure that the redundant versions running in parallel are independently designed and developed, e.g., by  different teams. Let us assume in particular that we are developing a fault-tolerant system that can tolerate up to one fault, for which case we need to use triple-redundancy~\cite{Laprie1990}. Hence we assume we have three programs $P_1$, $P_2$ and $P_3$ that have been developed by three independent teams from a same functional specification $Sp$ (see Figure~\ref{fig:firstExample}).

Before deploying them to the redundant architecture, we aim at ensuring that the three programs properly implement  $Sp$. Hence we also assume that a test suite $T_{0}$ has been derived from  $Sp$ using some specification-based approach.
The three independent teams will each test one program on the common 
$T_{0}$ test suite, and perhaps also derive additional test cases, but without any mutual cross-check.

Our approach instead proposes to monitor the coverage achieved by $T_{0}$ on  $P_1$, $P_2$ and $P_3$,  respectively. If we observe quite differing measures, we may wonder why three programs that should be in principle equivalent, when exposed to the same inputs achieve so different coverage measures.  Hence, we analyze the program that has achieved the lowest measure, say, e.g., $P_1$, and find some additional test cases  $T_1$ that can raise its coverage. We execute these same tests on $P_2$ and $P_3$, and analyze the results: if either of  them fails on the additional test cases, it probably  does not fully implement the specified functionality, e.g., it might lack  a proper check of the input variables. We can repeat this test suite augmentation process, finding further test cases that cover more unexecuted entities of the program that yields the lowest coverage so far, until we have all three programs adequately covered and no further failure is found.

In this example, to ensure that all three versions are adequately tested, we suggest that the coverage of one version is leveraged to augment the test suite of another version, and we expect that such approach could help finding missing functionalities.

\begin{figure}
	\centering
	 \includegraphics[width=1.0\columnwidth]{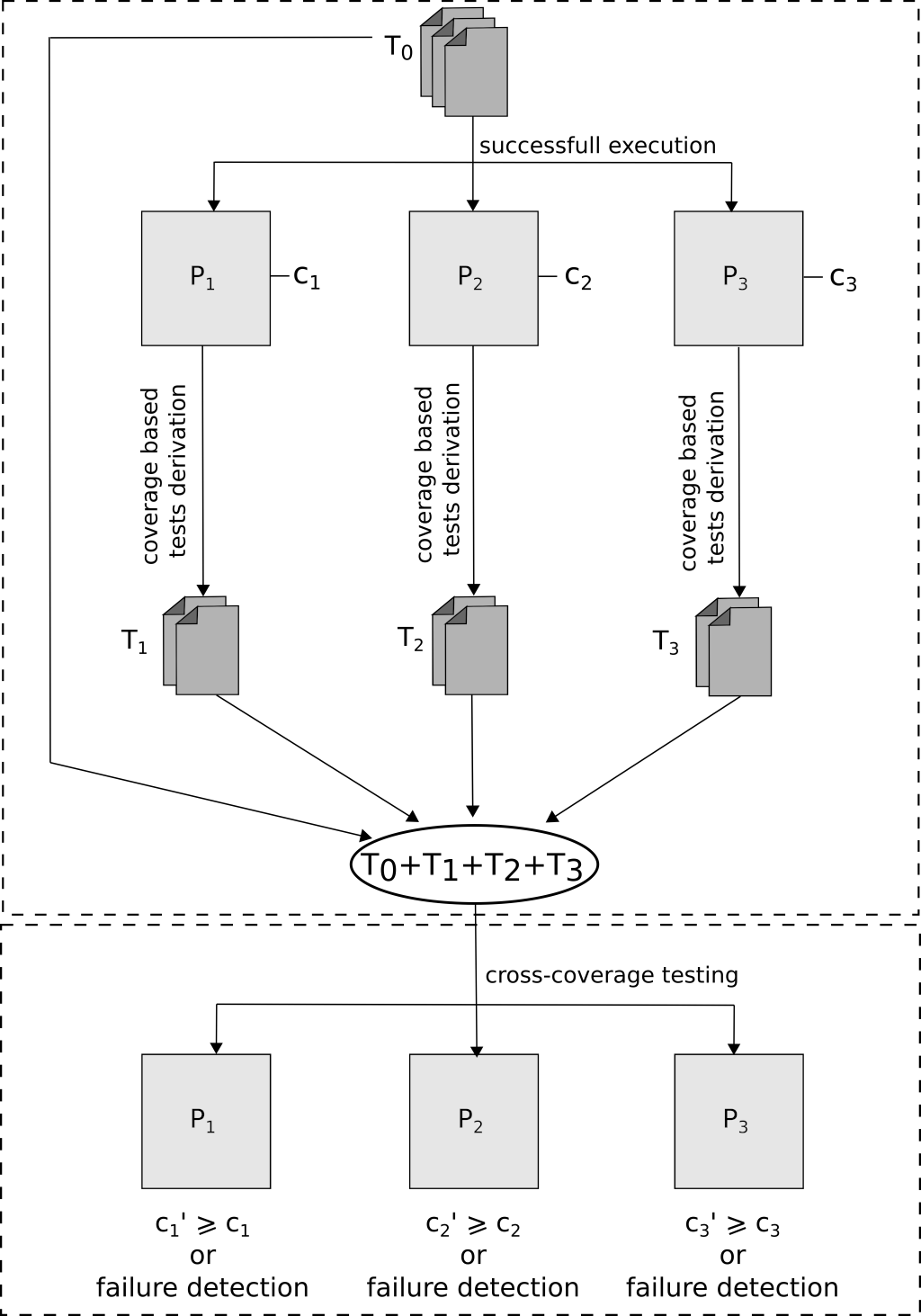}
	\caption{\label{fig:secondExample} \textit{Cross-coverage testing} for TDCS.}
\end{figure}

A  second motivating example (illustrated in Figure \ref{fig:secondExample}) concerns the reuse of existing code pieces in software development. Indeed, nowadays many different programs can be found in open-source code repositories such as GitHub, in web service directories, in app stores, as well as among proprietary solutions, for implementing one same desired function. As a matter of fact, several  studies~\cite{holmes2013systematizing,gharehyazie2017some,suzuki2017exploratory} confirm that developers increasingly perform code search to retrieve and reuse code snippets, methods or entire projects.  Among the many proposed approaches to make code search more effective and accurate~\cite{liu2021opportunities}, Lemos et al. have presented a Test-Driven Code Search (TDCS) approach~\cite{lemos2011test}. In brief, inspired by  test-driven  development,  they propose to first write a set of test cases that define the desired functionalities, and then to use such test cases both for searching the code repository for suitable matching implementations and for testing the found code pieces.
However the authors conclude the cited work by observing that in order
``\textit{to enhance confidence in the retrieved code, we could design test
cases in a more systematic way}''~\cite{lemos2011test}. Thus, they suggest that other testing criteria should be considered to integrate the initial design of test cases.

Our idea of cross-coverage can support  this recommendation.
Let us assume that thanks to TDCS we have found and retrieved three candidate programs $P_1$, $P_2$ and $P_3$ that all pass our initial test set $T_0$ (see Figure~\ref{fig:secondExample}). So far we have no further hint about which program to choose among  $P_1$, $P_2$ and $P_3$. We analyze the respective coverage achieved by the same test set $T_0$ over each of the three programs, which we denote as $c_1$,  $c_2$,  and $c_3$, respectively. For each program, we then define an additional set of test cases, $T_1$, $T_2$ and $T_3$ so to maximize its coverage (applying some traditional coverage-based test generation technique). Now we obtain an augmented test suite $T_{aug}$ by combining the initial test set $T_0$  with all the additional test cases, and execute this combined test suite on all three programs. Doing so we\textit{ (i) }may possibly  detect some failures and\textit{ (ii) } also obtain the increased coverage measures  $c'_1$,  $c'_2$,  and $c'_3$, respectively. We can now use $T_{aug}$ test results for choosing one program among $P_1$, $P_2$ and $P_3$; in particular we might prefer the one that  yields the highest coverage, because this method implements our desired functionalities with less unused code.

In both the above examples, the usage of cross-coverage information helps to augment the test suite in systematic and automatable way.
In the remainder of the paper, we formalize the notion of cross-coverage testing and provide a concrete evaluation of its usefulness.

\section{Cross-coverage testing}
\label{sec:approach}

In this section we present our cross-coverage testing approach and formally define the concepts of \emph{cross-coverage testing} and \emph{cross-coverage test suite augmentation} that are at the basis of our proposal.

First, for functional equivalence between programs, we rely on the same definition provided in~\cite{jiang2009automatic}, i.e., we say that two programs are \textit{functionally equivalent} if  given a same input,   they provide the same output, without considering the intermediate program states. Please refer to~\cite{jiang2009automatic} for further details. We can now introduce the concept of cross-coverage testing.

\begin{definsec}[Cross-coverage testing]
\label{crosscoverage}
Let $P_{1}, P_{2},\dots, P_{n}$
be \emph{n} functionally equivalent programs.
Given a test case $t_{i}$ derived to cover program $P_{i}$,
we define \emph{cross-coverage testing} as the approach of executing $t_{i}$ on $P_{j}$ with  $i,j= 1, \dots, n$ and $i\neq j$. The test case $t_{i}$ is said a \textit{cross-test} for  $P_{j}$, and the cross-coverage test execution is denoted as  $t_{i}\checkmark P_{j}$.
\end{definsec}

With the above definition we have introduced the concept of using coverage-driven test cases for testing a different program than the one from which they were derived. In other terms, by cross-coverage we mean in some sense that the test coverage for a program  $P_{j}$ could be monitored over another program $P_{i}$.

In the following, with reference to the schematic \textit{how-to} process outlined in the introduction, we focus on the usage of coverage for improving an initial test suite $T_{0}$, for example derived from the common specification (as in Figure \ref{fig:firstExample}).

\begin{definsec}[Cross-coverage test suite augmentation]
\label{augmentcrosstest}
Let $P_{1}, \dots, P_{n}$
be \emph{n}  functionally equivalent programs.
Let $T_{0}$ be a common initial test suite for $P_{1},\dots, P_{n}$.
Given a program $P_i$, let $T_{i}$ be a test suite of cross tests for $P_j$, with $i,j= 1, \dots, n$ and $i\neq j$.
\textit{Cross-coverage test suite augmentation} for $P_j$ consists in adding the tests in $T_{i}$ to  $T_{0}$, obtaining the \textit{augmented cross-test suite}  $(T_{0} \bigcup T_{i})$.
\end{definsec}

Referring to the above definition, we can have that equivalent programs $P_{1},\dots, P_{n}$ represent different implementations of the same functional specification. Then, we can suppose to have a common test suite $T_{0}$ for this set of equivalent programs, derived for instance from an abstract functional specification of these programs. Then, we can suppose to already have a specific test suite $T_{i}$  for the program $P_{i}$ derived to cover the specific implementation $P_{i}$. The main idea of the proposed approach is to leverage the coverage of the program $P_{i}$, and then the \emph{cross-tests}, namely the coverage-based test cases derived for $P_{i}$,  for augmenting the existing test suite $T_{0}$ with additional test cases aiming to  improve the testing of $P_{j}$.
The intuition behind our approach is that test cases generated to cover a program, for instance $P_{i}$, could have good chances to improve the testing of another program $P_{j}$ that is equivalent to $P_{i}$ and that implements the same functionalities of $P_{i}$.

In the above definition we purposely leave the characterization of a cross test quite generic.
Indeed, differing approaches could be taken for deriving the cross tests, for instance one could use tools such as EvoSuite~\footnote{\url{https://www.evosuite.org/}} or Klee~\footnote{\url{http://klee.github.io/}} to derive test cases that cover the not yet covered entities.

Note that this definition matches  the first motivating example in the previous section: looking to Figure~\ref{fig:firstExample}, the common initial test suite $T_{0}$ of the definition corresponds to the one derived from $Sp$,  and we augment it in iterative way, choosing at each iteration to apply the cross-coverage test suite augmentation approach by covering the program that has reached so far the least coverage.

While in the above example we consider that coverage of one program is used to drive the testing of all the other equivalent programs, we can also consider the complementary case that all equivalent  programs cumulatively   contribute to augment the test suite of a selected program, as in the following definition.

\begin{definsec}[Cumulative cross-coverage test suite augmentation]
\label{cumulativecrosstest}
Let $P_{1}, \dots, P_{n}$
be \emph{n}  functionally equivalent programs.
Let $T_{0}$ be a common initial test suite for $P_{1},\dots, P_{n}$.
Given a program $P_j$, let $T_{i}$ be a test suite of cross tests for $P_j$, with $i,j= 1, \dots, n$ and $i\neq j$.
\textit{Cumulative cross-coverage test suite augmentation}  for  $P_{j}$ consists in adding $\forall i \neq j$ the tests in $T_{i}$ to  $T_{0}$, obtaining the \textit{augmented cumulative cross-test suite}  $(T_{0} \bigcup \{T_{i}\})$.
\end{definsec}

The idea of the \textit{Cumulative cross-coverage test suite augmentation} of Definition \ref{cumulativecrosstest} is to consider all the \textit{augmented cross-test suite}s derived from all programs $P_i$ equivalent to $P_j$  and then leverage these test suites of all programs $P_i$ equivalent to $P_j$ for augmenting the initial test suite $T_{0}$.

Cumulative augmentation recalls  the case illustrated in the second motivating example (see Figure \ref{fig:secondExample}). However, note that in our example we have considered that one common test suite $T_{aug}$ is obtained as $(T_{0} \bigcup T_{1} \bigcup T_{2} \bigcup T_{3}  )$, i.e., it embeds all the additional test cases, whereas the above definition of the augmented cumulative cross-test suite would correspond to tree test suites, e.g.,  $(T_{0} \bigcup T_{2} \bigcup T_{3})$ when testing $P_1$, $(T_{0} \bigcup T_{1} \bigcup T_{3})$ when testing $P_2$, and $(T_{0} \bigcup T_{1} \bigcup T_{2})$ when testing $P_3$. In other terms, for this example we consider to test each program not only considering its cross tests but also the additional test cases that enhance its own coverage, which seems a sensible way to go when we can also monitor the coverage of the program considered.

The presented cross-coverage testing approach is not tailored to a specific test adequacy criterion \cite{zhu1997software}.
As we will show in Section \ref{sec:studyDesign}, for the aim of simplicity, in this paper we validated the proposed approach considering two coverage criteria that are: \emph{statement coverage} and \emph{branch coverage}, i.e, the  fraction of the total number of statements and the fraction of the total number of branches that have been exercised by a given test suite, respectively.
However, the proposed approach is generic and can address other common coverage measures including \emph{path coverage}, \emph{C-use}, \emph{P-use} coverage and others~\cite{frankl1988applicable}.

\section{Study Design}
\label{sec:studyDesign}

This section describes our study introducing the research questions (\rqs) that guided our validation (see Section~\ref{sec:rqs}), the subjects used for the empirical evaluation (see Section~\ref{sec:subjects}), and the methodology followed to answer the \rqs (see Section~\ref{sec:procedure}).

\subsection{Research Questions}
\label{sec:rqs}

Overall our study aims at evaluating whether \emph{cross-tests} contribute to improve the quality of a test suite when groups of functionally equivalent programs have to be tested. For addressing this general goal, we formulate two more specific research questions as follows.

\paragraph*{\rqOne}
Does cross-coverage test suite augmentation contribute to improve the thoroughness of a test suite? %How much?

\noindent By \rqOne we aim at assessing whether, given a program $P$, the additional cross tests that are derived based on the coverage of functionally equivalent alternatives, provide effective test cases also when executed on $P$.

In the scope of this study, as a proxy indicator for the effectiveness of a cross test we use the coverage it can achieve on the program under test, i.e., we observe if the cross tests actually help to explore parts of $P$ that have not yet been considered. Besides, our experiment will focus on cumulative cross-coverage test suite augmentation.

However, if the goal was just that of increasing the coverage of $P$,
we could as well, or even better, use one of the many existing code-based test generators directly on $P$, without having to resort to the coverage of other programs. In this sense, the usefulness of cross-coverage test suite augmentation as evaluated by \rqOne would be better appreciated in situations in which the code of $P$ is not available.
More in general, we foresee that cross-coverage can be useful also, or above all, to detect missing functionality faults among a group of supposedly equivalent programs. Thus, we formulate the second research question as follows.

\paragraph*{\rqTwo} Does cross-coverage test suite augmentation contribute to expose missing functionalities?

\noindent Overall, \rqTwo aims to study if cross-coverage testing is a valid approach for detecting undesired behaviour in a program $P$ due to missed functionalities in its implementation. Our intuition here is that the cross tests that are derived from the extensive coverage of a functionally equivalent program could also include test cases exercising functionalities that have been ``forgotten'' by the developers of $P$, and could certainly not be detectable using traditional coverage testing.

\subsection{Study subjects}
\label{sec:subjects}

We conducted our evaluation over 140 groups $G_k$, each one consisting of functionally equivalent programs (precisely Java methods), which have been selected from the large dataset built by Higo et al.~\cite{higo2022}\footnote{https://zenodo.org/record/5912689}.

In turn, the authors constructed this dataset from Borges et al.'s collection of GitHub projects~\cite{borges2016}: they first collected in groups those methods having the same return and parameter types, and derived for each method a set of coverage-based test cases using EvoSuite~\cite{fraser2011evosuite}. Then they mutually tested each method within a group using test cases automatically generated for all the other methods in the same group. Finally, all pairs of methods that passed each other test cases were subjected to visual inspection. As a result their dataset includes a total of 276 groups of functionally equivalent methods (728 methods in total).
The dataset provides an adequate benchmark for our study, because we get both groups of functionally equivalent programs and their test suites that are derived based on coverage (using EvoSuite on each method).

As each group refers to two or more equivalent programs exposing the same abstract functionality, we could also derive an additional test suite common to all the programs in each group. Specifically, we generated such a common
test suite using the feedback-directed random test generation approach by Randoop~\cite{PachecoLEB2007}. In the generation process we followed the typical configuration procedure as suggested by the Randoop user manual.

From the 276 available groups, in the scope of this experimentation we excluded those groups for which:
\begin{itemize}
 \item we were not able to generate the common test suite; or,
 \item for all the methods within the group, their specific test suite already included in Higo et al.'s benchmark
 could not be improved (i.e., the coverage already scored $100\%$).
\end{itemize}
Eventually, our benchmark consists of 140 groups including a total of
336 programs, their relative 336 test classes obtained from the benchmark in~\cite{higo2022} that are derived on the specific implementation of each program, plus 140 test classes that are common for each equivalent program in a group. Notably, while many groups only include two methods, there are also groups collecting 10 or more methods. An archive with all the data we referred to, the whole set of results, and the scripts for replicating our experiments is available at: \replicationpackage.

\subsection{Procedure and evaluation metrics}
\label{sec:procedure}

To answer \rqOne, we planned a two steps procedure, both steps referring to: a group of functionally equivalent programs $P_{1}, \dots, P_{n}$, their respective coverage-driven test suites $T_{1}, \dots, T_{n}$ generated using EvoSuite, and the common and implementation-independent test suite $T_{0}$ derived using Randoop.

In the first step, for each group we execute the common tests in $T_{0}$ on all programs of the group and measure for each program the coverage achieved ($T_{0}$ coverage). For coverage analysis, we use the tool JaCoCo and provide measures of statement and branch coverage.
Then, for each program $P_{i}$ in turn, we launch in an exhaustive way all the $T{j}$ (i.e.,  all the test suites provided in~\cite{higo2022} for the program $P_{j}$, with $i\neq j$): if for a fixed \textit{i} and for any \textit{j}, the cross test execution $t_{c}\checkmark P_{i}$ of at least one $t_{c}\in T{j}$ improves over the current coverage of $P_i$  (i.e., statements coverage or branches coverage is improved),
we iteratively add $T_{j}$ to the test suite for $P_{i}$. Eventually, at the end of this first step, for each program $P_i$ we obtain its augmented cumulative cross-test suite $CCT_{i} = (T_{0} \bigcup \{T_{j}\})$ and the final coverage achieved by this augmented test suite ($CCT$ coverage).
To answer \rqOne, we analyse the size of $CCT_{i}$ and the gains in coverage that it scores against the initial $T_{0}$ coverage.

As by construction the size of $CCT_{i}$ is greater than or equal to the size of $T_{0}$, it can appear as obvious that by adding more tests the coverage of $P_i$ increases. For a meaningful evaluation, it is important to assess the coverage gains against those that would be obtained by extending $T_{0}$ without taking into account any information on the programs equivalent to $P_{i}$. Thus, in the second step of the evaluation, for each $P_{i}$, we used again Randoop in order to extend $T_{0}$ to the same size of $CCT_{i}$ (denoted by $T^{+i}_{0}$); then we run $T^{+i}_{0}$ against $P_{i}$ and measure the coverage metrics it scores. The comparison of the coverage gains achieved by the baseline $T^{+i}_{0}$ against those from $CCT_{i}$ complements the answer to \rqOne.

To answer \rqTwo, we consider  all the $P_{i}$ recording some coverage improvement from \rqOne.
Among these,  chosen a program $P_{i}$, we
create a faulty version $MP_{i}$ where one of $P_{i}$'s original functionality has been dropped.
To do this, we
blindly and systematically modify $P_{i}$ by removing where possible either a branch of the first decision statement or one atom of its conditional expression. We verify that such modifications do not affect the compilation process. After that, for each $MP_{i}$ we run its respective $T_{0}$ (passing against $P_{i}$) and we remove all those $MP_{i}$ programs for which the modification is detected.

For the remaining $MP_{i}$, we separately run each $T_{j}$ against $MP_{i}$ counting each time the injected fault is spotted. The comparison between the times $MP_{i}$ have been detected as faulty by $T_{j}$ (but not $T_{0}$) against the whole set of cross-tests analysed gives an estimation of the efficacy in revealing missing functionalities of the cross-coverage testing across equivalent programs.

\section{Study Results}
\label{sec:studyResults}

In the following we report the outcome of the empirical evaluation and we provide answers to the \rqs given in Section~\ref{sec:rqs}.

\subsection{Capability to Improve  Test Suite Thoroughness}
\label{sec:answerRQ1}

In the first step of the empirical evaluation we employed $336$ programs from the benchmark in~\cite{higo2022}, as explained in Section ~\ref{sec:subjects}. On these we first applied  the general procedure planned to answer \rqOne (Section~\ref{sec:procedure}).

On each program $P_{i}^{k}$ within a group $G_k$,
we executed: the implementation-independent test suite we generated with Randoop and that is common to all the programs in a group (i.e., $T^{k}_{0}$);
plus all the  implementation-dependent test suites
$\{ T^{k}_{j} \}$, $j \neq i$,
available from the benchmark in~\cite{higo2022} and
generated using EvoSuite from the code of the other programs in the same equivalence group $G_k$ (i.e., when considering $P_i^{k}$ we did not execute the test suite $T^{k}_{i}$ EvoSuite generated from the code of $P_{i}^{k}$).

We provide below for each $P_{i}^{k}$ the measures achieved for coverage after applying cumulative cross-coverage test suite augmentation starting from the initial suite  $T^{k}_{0}$, and obtaining the augmented cumulative cross-test suite $CCT_{i}^{k}$. As described in the previous section, for simplicity in this study we add to the initial  test suite only those $T^{k}_{j}$ that actually bring some coverage increase.

Table~\ref{tab:effectiveness} reports the overall results of this first step of our experiment: across all the considered programs, in most of the cases (i.e. 86,90\%) we recorded that $P_{i}$ improved some of its coverage metrics when tested with $CCT_{i}$, while only for a minor part (i.e. 13,10\%) $CCT_{i}$ did not perform better than the initial implementation-independent test suite generated with Randoop (i.e, $T_{0}$).

\begin{table}
 \centering
 \begin{center}
\begin{tabular}{ccc}
Programs & Cases Without Augmentation & Cases With Augmentation\\
\hline
\hline
336 & 44 (13,10\%) & 292 (86,90\%)\\
\hline
 \end{tabular}
 \end{center}
\caption{Cumulative cross-coverage test suite augmentation}
\label{tab:effectiveness}
\end{table}

\begin{figure}
    \centering
    \begin{subfigure}[t]{0.9\columnwidth}
        \centering
        \includegraphics[width=\textwidth]{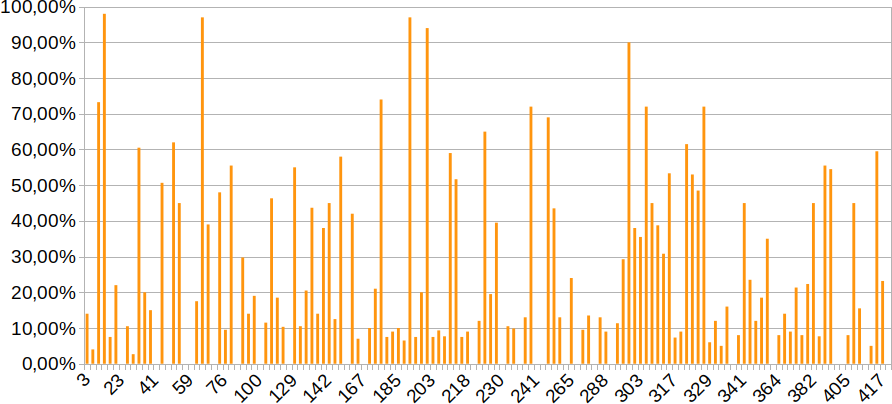}
        \caption{Statements}
        \label{fig:gains:statements}
    \end{subfigure}
    \quad
    \begin{subfigure}[t]{0.9\columnwidth}
        \centering
        \includegraphics[width=\textwidth]{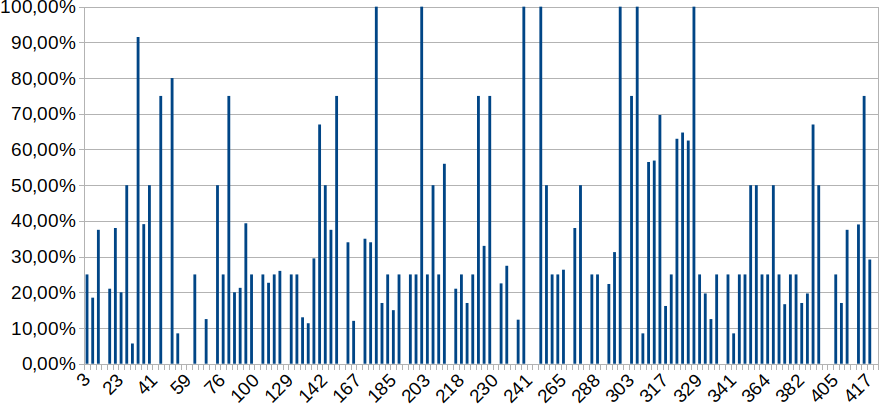}
        \caption{Branches}
        \label{fig:gains:branches}
    \end{subfigure}
    \caption{Average gain in coverage per groups: $CCT$ \emph{vs}. $T_{0}$}
    \label{fig:gains}
\end{figure}

Figure~\ref{fig:gains} further details such results by plotting the average gains
we measured in terms of both statements (Figure~\ref{fig:gains:statements}) and branches (Figure~\ref{fig:gains:branches}) coverage. The horizontal axis reports the group number (as referred in the original benchmark~\cite{higo2022}), while the vertical axis reports  the average percentage gain within a group; the  latter is  measured for each program $P_i^{k}$ within a group $G_k$ as the difference between the final percentage of own coverage achieved after executing $CCT_{i}^{k}$ and the initial percentage of own coverage achieved by $T^{k}_{0}$. Overall across the $336$ considered programs we register an average increase of $\sim26\%$ in statement coverage and of $\sim33\%$ in branch coverage.

Specifically, concerning statement coverage  (i.e., Figure~\ref{fig:gains:statements}), for $22$ groups we recorded an average gain between $[25-50\%)$, for $21$ groups the average gain in statements coverage is between $[50-75\%)$, for $5$ groups such average gain scores between $[75-100\%]$, while the other $66$ have an average gain
lower than $25\%$. For the remaining $26$ groups there is no benefit from the cumulative cross-coverage test augmentation process.

Similar results have been recorded for branch coverage (i.e., Figure~\ref{fig:gains:branches}): for $50$ groups the average gain is between $[25-50\%)$, $20$ groups score between $[50-75\%)$, $16$ groups have an average gain in the range $[75-100\%]$, while $28$ groups report
an increase lower than $25\%$. The remaining $26$ groups did not register any positive gain from the augmentation. Notably, among them $21$ groups overlap with those scoring $0.00\%$ gain in statements coverage, while in $10$ cases (i.e., $5+5$) we find a positive increase only in one of the two coverage metrics. Those groups that present a gain only on statements coverage refer to programs with no or few (e.g., only one) decisional statements where: some exception is explicitly thrown, the tests in $T^{k}_{0}$ force some unchecked exception, or a \texttt{return} statement is met before the last instruction of the program. The groups presenting only an increase in branches coverage again refer to simple programs with few decisional statements but with a complex boolean formula as conditional expression.

In Table~\ref{tab:gains} we provide an excerpt of more detailed results for a sample of seven groups (we anticipate these correspond to the groups that were identified in the second step of this experiment, as explained below),  whereas the detailed results for all the 336 programs can be found in the additional online material\footnote{\replicationpackage}.
The first and second column of the table report the Id of the groups and of the methods. Then we include two sets of three columns each, which report the specific coverage metrics (statements for columns 3-5, and branches for columns 6-8) scored for each program by  $CCT_{i}^{k}$ and $T^{k}_{0}$, and their difference (i.e., Gain). Also, column 9 shows the size of the test suite resulting from the cross-coverage augmentation procedure (i.e., $T^{k}_{0}$ plus some $T^{k}_{j}$ with $j \neq i$): we see that  in most cases $CCT_{i}^{k}$ adds to $T^{k}_{0}$ only one implementation-dependent test suite $T^{k}_{j}$ from some program equivalent to $P_{i}^{k}$. Overall the average $CCT$ size for those test programs that increase some coverage metric is $2,05$ test suites.
In our intuition this result was somehow expected for the considered benchmark: the tests in~\cite{higo2022} were generated by means of EvoSuite which (within the given budget) looks for specific configurations maximising the coverage of the considered program. Indeed, the test in $T^{k}_{j}$ are selected so that to force the activation of several branches and conditions in $P_{j}^{k}$ with the objective of highlighting multiple usage scenarios for the referred functionality. In addition, the programs in the benchmark are relatively small, and the scenarios explored by $T^{k}_{j}$ could overlap with those covered by other test suites in $G_{k}$. Thus, limited to this benchmark, we found reasonable that on average a bit more than one additional test suite (i.e., $CCT_{i}^{k} = T^{k}_{0} \bigcup \{T^{k}_{j}\}$ for $j\neq i$) is the best solution for $P_{i}^{k}$.

While these  results are encouraging, we also assessed what coverage gains could be scored by a test suite having the same size as $CCT_{i}^{k}$ but generated without taking into account any cross-coverage information.
We hence took as a baseline an augmented test suite $T^{k+i}_{0}$ obtained from $T^{k}_{0}$ by adding the proper set of additional random test cases (which depends on the resulting size of $CCT_{i}^{k}$), generated by using again Randoop.
For this purpose we selected all the groups that from the previous step got a resulting average score in statements coverage lower than $85\%$.
The rationale was to give higher chances to the random baseline to beat  our approach (because it is well known that the higher is the current coverage the more difficult is  to find test cases that can increase it).
Such first selection returned with $67$  programs. From them we excluded all those programs (i.e., $28$) with no actual augmentation (i.e., $CCT_{i}^{k} = T^{k}_{0}$), and then we also removed all those programs (i.e., $17$) where the typical configuration procedure suggested by the Randoop user manual produces a test suite lower in size than $CCT_{i}^{k}$.
The resulting set of equivalent test programs (i.e., $22$) is reported in Table~\ref{tab:baseline}\footnote{As anticipated, the excerpt reported in Table~\ref{tab:gains} refers to the same set of equivalent programs.}.  In the third and fourth columns we report for each program  the difference between the increase  of statement and branch coverage achieved by $CCT$ and $T^{+}_{0}$. Then, columns 5 and 6 provide the average values of such differences.

 \begin{table*}
 \centering
 \begin{center}
\begin{tabular}{cc|p{0.06\textwidth}p{0.06\textwidth}p{0.06\textwidth}|p{0.06\textwidth}p{0.06\textwidth}p{0.06\textwidth}|c|p{0.06\textwidth}p{0.07\textwidth}p{0.06\textwidth}}
\multirow{2}{*}{GroupId} & \multirow{2}{*}{MethodId} & \multicolumn{3}{c}{STMS} & \multicolumn{3}{c}{BRANCHES} & $CCT$ & \multicolumn{3}{c}{Group AVG} \\
 &  & $CCT$ & $T_{0}$ & Gain & $CCT$ & $T_{0}$ & Gain & Size & Gain STMS & Gain BRANCHES & $CCT$ Size\\
\hline                       
\hline
\multirow{12}{*}{37}   & 10881  & 91,00\%  & 65,00\% & 26,00\%	& 87,00\%  & 50,00\% & 	37,00\% & 2 & \multirow{12}{0.04\textwidth}{20,00\%} & \multirow{12}{0.04\textwidth}{39,08\%} & \multirow{12}{0.04\textwidth}{2,25}\\
                       & 54076  & 17,00\%  & 15,00\% & 2,00\%	& 16,00\%  & 12,00\% & 	4,00\%  & 2 &  &  &  \\
                       & 61760  & 100,00\% & 78,00\% & 22,00\%	& 100,00\% & 50,00\% & 	50,00\% & 2 &  &  &  \\
                       & 67630  & 29,00\%  & 28,00\% & 1,00\%	& 32,00\%  & 26,00\% & 	6,00\%  & 3 &  &  &  \\
                       & 101187 & 76,00\%  & 61,00\% & 15,00\%	& 83,00\%  & 50,00\% & 	33,00\% & 2 &  &  &  \\
                       & 181606 & 100,00\% & 78,00\% & 22,00\%	& 100,00\% & 50,00\% & 	50,00\% & 2 &  &  &  \\
                       & 181644 & 100,00\% & 88,00\% & 12,00\%	& 100,00\% & 66,00\% & 	34,00\% & 2 &  &  &  \\
                       & 184888 & 100,00\% & 60,00\% & 40,00\%	& 100,00\% & 25,00\% & 	75,00\% & 2 &  &  &  \\
                       & 191816 & 100,00\% & 55,00\% & 45,00\%	& 100,00\% & 37,00\% & 	63,00\% & 3 &  &  &  \\
                       & 277587 & 100,00\% & 68,00\% & 32,00\%	& 100,00\% & 50,00\% & 	50,00\% & 3 &  &  &  \\
                       & 296201 & 88,00\%  & 77,00\% & 11,00\%	& 83,00\%  & 50,00\% & 	33,00\% & 2 &  &  &  \\
                       & 296211 & 100,00\% & 88,00\% & 12,00\%	& 100,00\% & 66,00\% & 	34,00\% & 2 &  &  &  \\
\hline                                                                                              
\multirow{2}{*}{40}    & 267714 & 76,00\%  & 61,00\% & 15,00\%	& 100,00\% & 50,00\% & 	50,00\%	& 2 & \multirow{2}{0.04\textwidth}{15,00\%} & \multirow{2}{0.04\textwidth}{50,00\%} & \multirow{2}{0.04\textwidth}{2,00} \\
                       & 292418 & 76,00\%  & 61,00\% & 15,00\%	& 100,00\% & 50,00\% & 	50,00\%	& 2 &  &  &  \\
\hline                                                                                              
\multirow{2}{*}{80}    & 242345 & 69,00\%  & 23,00\% & 46,00\%	& 62,00\%  & 12,00\% & 	50,00\%	& 2 & \multirow{2}{0.04\textwidth}{55,50\%} & \multirow{2}{0.04\textwidth}{75,00\%} & \multirow{2}{0.04\textwidth}{2,00}\\
                       & 242357 & 100,00\% & 35,00\% & 65,00\%	& 100,00\% & 0,00\% & 	100,00\%& 2 &  &  &  \\
\hline                                                                                              
\multirow{2}{*}{106}   & 225177 & 65,00\%  & 65,00\% & 0,00\%	& 50,00\%  & 50,00\% & 	0,00\%	& 1 & \multirow{2}{*}{11,50\%} & \multirow{2}{*}{25,00\%} & \multirow{2}{*}{1,50}\\
                       & 250021 & 100,00\% & 77,00\% & 23,00\%	& 100,00\% & 50,00\% & 	50,00\%	& 2 &  &  &  \\
\hline                                                                                              
\multirow{2}{*}{225}   & 137547 & 81,00\%  & 62,00\% & 19,00\%	& 66,00\%  & 33,00\% & 	33,00\%	& 2 & \multirow{2}{0.04\textwidth}{19,50\%} & \multirow{2}{0.04\textwidth}{33,00\%} & \multirow{2}{0.04\textwidth}{2,00} \\
                       & 154527 & 80,00\%  & 60,00\% & 20,00\%	& 66,00\%  & 33,00\% & 	33,00\%	& 2 &  &  &  \\
\hline                                                                                              
\multirow{2}{*}{343}   & 20346  & 100,00\% & 53,00\% & 47,00\%	& 100,00\% & 50,00\% & 	50,00\%	& 2 & \multirow{2}{*}{23,50\%} & \multirow{2}{*}{0,50\%} & \multirow{2}{*}{1,50} \\
                       & 251573 & 63,00\%  & 63,00\% & 0,00\%	& 50,00\%  & 50,00\% & 	0,00\%	& 1 &  &  &  \\
\hline                                                                                              
\multirow{3}{*}{369}   & 77210  & 46,00\%  & 46,00\% & 0,00\%	& 33,00\%  & 33,00\% & 	0,00\%	& 1 & \multirow{3}{0.04\textwidth}{21,33\%} & \multirow{3}{0.04\textwidth}{16,67\%} & \multirow{3}{0.04\textwidth}{2,00} \\
                       & 102898 & 100,00\% & 70,00\% & 30,00\%	& 100,00\% & 50,00\% & 	50,00\%	& 3 &  &  &  \\
                       & 114340 & 100,00\% & 66,00\% & 34,00\%	& 100,00\% & 100,00\%& 	0,00\%	& 2 &  &  &  \\
\hline                       
 \end{tabular}
 \end{center}
 \caption{Thoroughness: Excerpt of few gains from the cross-coverage test suite augmentation}
 \label{tab:gains}
\end{table*}

As shown, the results from the comparison between the cumulative cross-coverage test suite augmentation process and the baseline clearly confirm that cross-coverage positively contributes to improve a test suite thoroughness, better than random generation. From the table, we see only one case where random augmentation scores higher coverage than cross-coverage (program 67630 of Group 37). In all the other cases $CCT$ reaches always higher or much higher coverage than the baseline. Specifically, for these program cross-tests achieve an average coverage increase that is $\sim21\%$ for statement and $\sim22\%$ for branch more than the increase obtained by the baseline.

Thus we can answer \rqOne by reporting one of the outcome from our study: implementation-dependent tests from equivalent programs contribute to improve the test suites thoroughness. Moreover, the gain from such cross-coverage process is more effective than a random augmentation of the numbers of the tests.

 \begin{table}
 \centering
 \begin{center}
\begin{tabular}{cc|p{0.11\columnwidth}p{0.11\columnwidth}|p{0.08\columnwidth}p{0.08\columnwidth}}
\multirow{2}{*}{GroupId} & \multirow{2}{*}{MethodId} & \multicolumn{2}{c}{Gain: {\footnotesize ($CCT - T^{+}_{0}$)}} & \multicolumn{2}{c}{Group AVG} \\
 &  & STMS & BRNS & STMS & BRNS \\
\hline
\hline
\multirow{12}{*}{37}   & 10881  & 26,00\%  & 37,00\%  & \multirow{12}{0.04\columnwidth}{11,58\%} & \multirow{12}{0.04\columnwidth}{19,83\%} \\
                       & 54076  & 2,00\%   & 4,00\%   &  &  \\
                       & 61760  & 11,00\%  & 13,00\%  &  &  \\
                       & 67630  & -10,00\% & -10,00\% &  &  \\
                       & 101187 & 7,00\%   & 17,00\%  &  &  \\
                       & 181606 & 11,00\%  & 13,00\%  &  &  \\
                       & 181644 & 12,00\%  & 17,00\%  &  &  \\
                       & 184888 & 14,00\%  & 25,00\%  &  &  \\
                       & 191816 & 32,00\%  & 38,00\%  &  &  \\
                       & 277587 & 11,00\%  & 17,00\%  &  &  \\
                       & 296201 & 11,00\%  & 33,00\%  &  &  \\
                       & 296211 & 12,00\%  & 34,00\%  &  &  \\
\hline
\multirow{2}{*}{40}    & 267714 & 15,00\%  & 50,00\%  & \multirow{2}{0.04\columnwidth}{15,00\%} & \multirow{2}{0.04\columnwidth}{50,00\%}\\
                       & 292418 & 15,00\%  & 50,00\%  &  &  \\
\hline
\multirow{2}{*}{80}    & 242345 & 46,00\%  & 50,00\%  & \multirow{2}{0.04\columnwidth}{55,50\%} & \multirow{2}{0.04\columnwidth}{75,00\%} \\
                       & 242357 & 65,00\%  & 100,00\% &  &  \\
\hline
106                    & 250021 & 23,00\%  & 50,00\%  & 23,00\% & 50,00\% \\
\hline
\multirow{2}{*}{225}   & 137547 & 19,00\%  & 33,00\%  & \multirow{2}{0.04\columnwidth}{19,50\%} & \multirow{2}{0.04\columnwidth}{33,00\%} \\
                       & 154527 & 20,00\%  & 33,00\%  &  &  \\
\hline
343                    & 20346  & 47,00\%  & 50,00\%  & 47,00\% & 50,00\% \\
\hline
\multirow{2}{*}{369}   & 102898 & 30,00\%  & 50,00\%  & \multirow{2}{0.04\columnwidth}{32,00\%} & \multirow{2}{0.04\columnwidth}{25,00\%} \\
                       & 114340 & 34,00\%  & 0,00\%   &  &  \\
\hline
 \end{tabular}
 \end{center}
 \caption{Comparison with the baseline: $CCT$ \emph{vs}. $T^{+}_{0}$.}
 \label{tab:baseline}
\end{table}

\subsection{Capability to Detect Missing Functionalities}
\label{sec:answerRQ2}

About \rqTwo, we referred all the $292$  programs $P_{i}$ resulting from the first validation step we performed for \rqOne (see Table \ref{tab:effectiveness}). As introduced in Section~\ref{sec:procedure}, for each of these $P_{i}$ we
considered:
$MP_{i}$, which is a faulty version of $P_{i}$ where one of its functionality has been removed; the initial  (implementation-independent) test suite $T_{0}$; and the set of
$T_{j}$ (with $i\neq j$), i.e., the implementation-dependent test suites of the programs equivalent to $P_{i}$.
After removing all $MP_{i}$ whose modification is detected by the initial test suite $T_{0}$, we got $187$ modified programs.
We  compared these remaining $MP_{i}$ against $P_{i}$ to double-check the applied modification is actually subsuming a missed functionality: we further excluded $18$ $MP_{i}$. Overall we are left with $169$ valid modified versions of the original programs. For each of them, within their respective group $G_k$ we separately considered all the $T^{k}_{j}$ in the $CCT_{i}^{k}$ resulting from \rqOne and added to $T^{k}_{0}$. As some equivalent test programs have multiple items in $CCT_{i}^{k}$, the final set of valid modified programs undetected by $T^{k}_{0}$ to be considered for the cross-check analysis counts $183$ items.

\begin{table}
 \centering
 \begin{center}
\begin{tabular}{l|c}
\hline
Modified Programs Undetected by $T_{0}$ & 187\\
\hline
Invalid Modifications & 18\\
\hline
Valid Modified Programs & 169\\
\hline
Valid Cases For Cross-Check & 183\\
\hline
Valid Cases Detected by CCT & 132 (72,13\%)\\
\hline
Valid Cases Undetected by CCT & 51 (27,87\%) \\
\hline
Valid Modified Programs Detected by CCT & 124 (73,37\%)\\
\hline
Valid Modified Programs Undetected by CCT & 45 (26,63\%)\\
\hline
 \end{tabular}

 \end{center}
\caption{Results on detection of missing functionalities}
\label{tab:efficacy}
\end{table}

Table~\ref{tab:efficacy} summarises the results of this second study: for $132$ items the missing functionality is revealed by at least one of the implementation-dependent test suites of equivalent programs, while for the remaining $51$ items, no specific implementation-dependent test suite of the programs equivalent to $P_{i}$ is able to detect the missed functionality. More in detail about these last cases, the results report that for cases $2$ the missing functionality is actually revealed when the whole $CCT_{i}^{k}$ is run, thus the actual number of programs for which the modification is detected counts $124$, while the modifications are never detected for $45$ programs.

In conclusion, also our answer to \rqTwo is positive: cross-coverage testing from equivalent programs is a valuable strategy for exposing issues related to missing functionalities. The outcomes from our study show that, over $274$ (i.e., $292-18$) programs, the implementation-dependent tests from equivalent programs help to detect missing functionalities for $73,37\%$ out of the $169$ cases in which the initial implementation-independent test suites do not spot them.

\section{Threats to validity}
\label{sec:threats}

Even though the outcomes from our study lead to positive answers for both the \rqs, we highlight some threats that may potentially affect the validity of our empirical evaluation and the conclusions we have drawn.

\subsection{Threats to External Validity}
\label{sec:threats:extenal}

This class of threats to validity address those concerns limiting the generalisation of the observed results to other studies or contexts.

\paragraph*{Considered Subject} A first class of concerns come from the referred benchmark by Higo et al.~\cite{higo2022}. Indeed, the programs in~\cite{higo2022} are actually methods extracted from wider Java classes, thus: their complexity is relatively limited, and they implement state-less functions only. Our work does not cover yet the study of complex software where the implementations of system-level features require the interaction across several (possibly state-full) software bundles. In this sense, even thought we are confident that tests from specific usage scenario of a whole system could be also valuable for checking missing functionalities in equivalent software, our conclusions do not have a general validity for any kind of software system.

\paragraph*{Functional Equivalence} We are aware that the definition of functional equivalence referred from~\cite{jiang2009automatic} and also used in~\cite{higo2022} may not be accepted in all the contexts. In this sense our conclusions are strictly related to such definition, and they cannot be referred elsewhere if such definition does not hold.

\subsection{Threats to Internal Validity}
\label{sec:threats:internal}

This class of threats to validity refers to those aspects that could have influenced the observed outcome.

\paragraph*{Generation of $T_{0}$}
As reported in Section~\ref{sec:subjects}, for the generation of the common implementation-independent test suites we followed the typical configuration procedure reported by the Randoop user manual. We are aware that we used only the basic functionalities by Randoop. Thus, we do not exclude that a deeper tuning of its parameters may have led to the generation of more powerful $T_{0}$ for each of the considered programs. Probably in this case our experiment would have been resulted with different outcomes. However, our decision is motivated by both our limited confidence in tuning Randoop, and also the limited complexity of the programs in the subject.

\paragraph*{Construction of $CCT$}
About the cumulative construction of the cross-coverage test suite, for each program  $P_{i}$ we randomly selected one of its equivalent program $P_{j}$ (with $i\neq j$) and we launched its $T_{j}$ against $P_{i}$: if the cross testing improves the current coverage metrics, then $T_{j}$ is added to $CCT_{i}$. Indeed, the specific order the equivalent programs $P_{j}$ are selected can impact the construction of $CCT_{i}$: the resulting $CCT_{i}$ cannot be considered the only admissible solution for $P_{i}$. Thus, some of the indexes presented in Table~\ref{tab:gains} (e.g., $CTT$ Size), in Table~\ref{tab:efficacy} (e.g., Valid Cases For Cross-Check), or discussed in Section~\ref{sec:studyResults} may result with small but different values across next replications of the experiment.

\paragraph*{Construction of $T_{0}$ and $T^{+}_{0}$}
Differently from the construction of $CCT$, all the tests inferred by Randoop were then included either in $T_{0}$ or $T^{+}_{0}$. Indeed, these tests are independent from the specific implementations of the $P_{i}$ and they could not be compared in terms of any structural metric. Thus, we added each test to $T_{0}$ or to $T^{+}_{0}$ without checking its actual quality (e.g., expressed as an improvement of any coverage metric). We cannot exclude that this procedure did not affect our study.

\paragraph*{Programs modification}
In the experiment that answers to \rqTwo, a potential threat to validity could also come from the specific procedure followed for creating the modified version of each $P_{i}$. Indeed, in order to get the faulty version $MP_{i}$ we first agreed on a common strategy, without taking into account the actual implementation of $P_{i}$; then we blindly applied it on the code base. We did our best to avoid biases in both the selection of the strategy, and its application; nevertheless we cannot exclude that our procedure did not affect any outcome or our conclusions.

\section{Related work}
\label{sec:related}
The related work of our proposal spans over two main research areas that are: testing of equivalent programs and test suite augmentation.

\paragraph{Testing of equivalent programs}
Code fragments having functionally equivalent behavior, known also as \emph{software redundancy}, are largely used in software development, for fault tolerance purposes (as in the case of \emph{N-version} programming \cite{avizienis1985n}) but also in  self-healing systems \cite{carzaniga2013automatic} or self-adaptive services \cite{sikri2014adaptive}.
The problem of finding or checking semantically equivalent code has been largely investigated in the literature since decades \cite{cousineau1979program}. Some of the existing studies  focus on software testing for checking functional equivalence of two programs. For instance,
the work in \cite{jakobs2022peqtest} aims to check functional equivalence of refactored code segments by generating test programs to detect variable values inequivalence whereas that in \cite{jiang2009automatic} applies random testing on arbitrary pieces of code
and detects their equivalence if they show the same output behavior  with the same input.
Other studies address the problem of measuring the redundant software in order to predict the effectiveness of different redundancy techniques \cite{carzaniga2015measuring}.
The goal of this paper is totally different from such previous studies. Our goal is to provide a new testing approach that we call \textit{cross-coverage testing} to improve the testing of two or more programs that are (assumed to be) functionally equivalent.

Few studies address the problem of effective testing of equivalent or redundant programs.
In fault tolerant systems, \emph{back-to-back testing} is used to check the outputs obtained from functionally equivalent versions of the same program. If the outputs are equal, the versions are correct, if they are different, further investigations are made to locate faults. In \emph{back-to-back testing}, test cases are usually randomly generated and include boundary or special values. Besides \emph{back-to-back testing} is unable to detect identical incorrect outputs \cite{brilliant1990performance}, it has been proven to be effective in detecting correlated faults \cite{vouk1988back}.
The authors of \cite{lyu1994coverage} provide the ATAC coverage analysis tool and show its application to detect a correlation between the number of faults and the coverage of 12 program versions of a critical application developed into a \emph{N-version} software project. The tool also allows to generate new tests aiming to improve coverage by examining not covered code.
In our approach the aim is not to improve the test suite according to the uncovered code of a given program under test but to leverage the mutual coverage analysis of existing test suites derived from equivalent versions of a program for improving the testing of the overall redundant system.

\paragraph{Test suite augmentation}

Existing test suites augmentation proposals aim to identify  the code elements affected by changes, typically during regression testing, and then generate test cases to cover those elements \cite{xu2010directed}.
These solutions rely on genetic algorithms \cite{xu2010factors}, symbolic execution \cite{santelices2008test} or the combination of different test cases generation algorithms \cite{xu2011hybrid} for augmenting the test suite.
Other approaches proposed in literature for test suite augmentation leverage search based techniques for test data regeneration starting from pre-existing test cases \cite{yoo2012test} or aim to derive integration tests from existing unit tests \cite{pezze2013generating}.
The idea of improving a test suite  by adding variants of existing test cases (for instance manually written by developers) is also known as test amplification \cite{danglot2019snowballing}. The concept of test amplification is a general concept that includes test augmentation.
Along with mutation analysis \cite{smith2009guiding}, coverage analysis represents one of the main  engineering goals guiding test amplification \cite{tzoref2022tackletest} and test augmentation \cite{bloem2014automating}.
Coverage analysis has been used for decades for different purposes, i.e., for providing an indication of the thoroughness of the executed test cases \cite{inozemtseva2014coverage},  for test case prioritization \cite{khatibsyarbini2018test}, for prediction of software reliability \cite{963124} or for  selection of
effective test cases \cite{bertolino2019adaptive}.
In test augmentation, given an initial test suite, additional test cases are added in order to  maximize the coverage of the program code.
However, most existing test suite augmentation techniques generate new test cases guided  by the coverage analysis of the very program under test.
In our approach, instead, the generation of new test cases is guided  by  the coverage analysis of a set of functionally equivalent programs.

Indeed, few approaches address test augmentation considering multiple versions of a program as in our proposal. For instance, Osei-Owusu et al. \cite{osei2019grading} perform test suite augmentation for assessing the correctness of  multiple implementations of the same specification. Additional test cases are derived such that they are able to detect incorrect programs according to a defined concept of behavioral-equivalence approximation.
Gaston and Clause \cite{gaston2020method} propose a statical analysis technique that allows to identify missing unit tests by analysing sibling groups, i.e. groups of methods in different applications that
accomplish the same goal in different ways, and their respective test suites defined in Junit.
However, these works do not focus on coverage analysis of equivalent programs to identify additional tests.
Moreover, Kessel and Atkinson \cite{kessel2019platform} present a general framework leveraging redundant implementations of the software under test for: i) test amplification by reusing existing tests for alternative implementations of a functional abstraction; ii) test output estimation, for instance taking the responses of the majority of redundant implementations; and finally iii) test suite evaluation, by enhancing the tests set with tests that raised a discrepancy in the responses  of the redundant implementations. On top of this  general framework, Kessel and Atkinson  \cite{kessel2022diversity} also provide an automated solution for diversity-driven tests generation. They apply EvoSuite to a set of alternative implementations  and compare the performance of the derived test suite with that achieved by the EvoSuite application to a single implementation in terms of mutation and branch coverage.
As in the previously cited works of Kessel and Atkinson~\cite{kessel2019platform,kessel2022diversity}, our approach also  leverages the test suites of alternative implementations for test suite augmentation.
However, differently from~\cite{kessel2019platform,kessel2022diversity}, our ultimate goal is that of leveraging the test cases derived from the coverage of other programs  to improve the test suite of a given program or to improve cost-effectiveness when testing redundant software. Moreover,  we present here a formal definition of cross-coverage, which to the best of our knowledge has never been  introduced before.

\section{Conclusion and Future Work}
\label{sec:conclusions}

Although largely investigated in the literature, research related to coverage analysis still attracts interest within the software engineering community. In this paper, we defined the new  concept of \emph{cross-coverage testing} of equivalent programs. The main idea is to leverage the coverage driven test cases derived for a program to improve the testing of another program that is equivalent to the first one.
Building on the foundational concept of \emph{cross-coverage testing} we defined \emph{cross coverage test suite augmentation} and  \emph{cumulative cross coverage test suite augmentation}  concepts, discussing their application in two representative contexts dealing with N-version programming and test-driven code search for software reuse, respectively.
Experimental results on a benchmark of 336 real programs collected into 140 functionally equivalent groups  have demonstrated the effectiveness of \emph{cumulative cross coverage test suite augmentation}, as cross tests reached an average increase of $\sim26\%$ in statement coverage and $\sim33\%$  in branch coverage.

Moreover, for a subset of 22 programs, selected among those for which random tests have higher chances to increase coverage (i.e. those showing average statement coverage lower than $85\%$), cross tests showed an average improvement of $\sim21\%$ and $\sim22\%$ of statement and branch coverage respectively, with respect to randomly generated test suites with the same cardinality, chosen as a baseline.

In our experiment we also showed the ability of the augmented test suite to discover $\sim73\%$ of missing faults introduced by systematically modifying test programs.
Therefore, the proposed cross coverage testing brings gains both in thoroughness  of a test suite and in exposing missing functionalities.

Future work is foreseen in several directions. A set of more extensive experiments based on larger benchmark programs is a first planned extension.
We also want to compare the effectiveness of our proposals  of \emph{cross coverage test suite augmentation} and \emph{cumulative cross coverage test suite augmentation}
with other existing test augmentation approaches, based for instance on mutation analysis.
In addition, cross coverage testing application in the two representative contexts considered in the paper  would deserve deeper investigation, considering for instance a real redundancy based fault tolerant system.

From the perspective of demonstrating the potential generality of cross coverage testing approach and its usefulness in a variety of applications, we have already identified the usage of this technique for  guiding the systematic  testing for a program whose code is not available.
For instance, cross coverage testing could be applied to test multiple programs when the code of one or more of them is proprietary (only the specification could be available), meaning that their coverage analyzer is not available. In this case, cross tests could be useful for improving black-box testing of these programs.

\section*{Acknowledgment}
This work has been partially supported by \sismaproject, and \indamgncsproject.
The authors wish to thank the anonymous AST reviewers for their insightful comments.

\balance
\bibliographystyle{IEEEtran}
\bibliography{biblioSlimVersion}

\end{document}